 \documentclass[final,5p,times,twocolumn,authoryear]{elsarticle}

\usepackage{amsmath}
\usepackage{amssymb}
\usepackage{lipsum}
\usepackage{natbib}
\usepackage{hyperref}
\usepackage[T1]{fontenc}

\journal{Astronomy $\&$ Computing}

\begin{document}

\begin{frontmatter}



\title{Advancements in Glitch Subtraction Systems for Enhancing Gravitational Wave Data Analysis: A Brief Review}


\author[first,second]{Mohammad Abu Thaher Chowdhury}
\affiliation[first]{organization={Department of Physics and Astronomy, University of Texas Rio Grande Valley},
            addressline={One West University Blvd.,}, 
            city={Brownsville},
            postcode={78520}, 
            state={Texas},
            country={USA}}
\affiliation[second]{organization={Department of Physics, Applied Physics and Astronomy, Rensselaer Polytechnic Institute},
            addressline={110 8th st,}, 
            city={Troy},
            postcode={12180}, 
            state={New York},
            country={USA}}

\begin{abstract}
Glitches are transitory noise artifacts that degrade the detection sensitivity and accuracy of interferometric observatories such as LIGO and Virgo in gravitational wave astronomy. Reliable glitch subtraction techniques are essential for separating genuine gravitational wave signals from background noise and improving the accuracy of astrophysical investigations. This review study summarizes the some of the main glitch subtraction methods used in the industry. We talk about the efficacy of classic time-domain techniques in real-time applications, like matched filtering and regression methods. The robustness of frequency-domain approaches, such as wavelet transformations and spectral analysis, in detecting and mitigating non-stationary glitches is assessed. We also investigate sophisticated machine learning methods, demonstrating great potential in automatically identifying and eliminating intricate glitch patterns. We hope to provide a thorough understanding of these approaches' uses, difficulties, and potential for future development in gravitational wave data analysis by contrasting their advantages and disadvantages. Researchers looking to enhance glitch subtraction procedures and raise the accuracy of gravitational wave detections will find great value in this paper.
\end{abstract}



\begin{keyword}
Glitch \sep Gravitational wave data \sep Computational methods \sep Multi-messenger Astronomy 



\end{keyword}

\end{frontmatter}




\section{Introduction}
\label{introduction}

Gravitational wave astronomy has ushered in a new era of multimessanger astronomy, providing unprecedented insights into some of the universe's most enigmatic phenomena. However, the accurate detection and characterization of gravitational wave signals amidst instrumental noise, known as "glitches," pose significant challenges~\citep{Abott_2016}. Glitches arise from various sources, including instrumental artifacts, environmental disturbances, and transient astrophysical events, obscuring genuine gravitational wave signals and impeding scientific interpretation.

Over the past decades, the development of advanced glitch removal systems has emerged as a critical area of research within the gravitational wave astronomy community. These systems employ diverse algorithms and techniques to identify and mitigate glitches to enhance the sensitivity and reliability of gravitational wave detectors. From traditional threshold-based approaches to sophisticated machine learning methods, the field has witnessed remarkable progress in improving the quality of gravitational wave data through effective glitch mitigation strategies.

This paper comprehensively reviews the state-of-the-art glitch removal systems deployed in gravitational wave data. We systematically survey the key methodologies and algorithms employed in glitch identification and removal, highlighting their strengths, and limitations. Notable algorithms include the Time-Frequency Wavelet Denoising~\citep{Cornish2020}, the BayesWave algorithm~\citep{Cornish_2015}, machine learning techniques such as convolutional neural networks~\citep{GEORGE2018}, gwsubtract~\citep{Davis_2019}, glitschen~\citep{Merritt2021}, deepclean~\citep{Ormiston_2020}, Swarm Heuristics based Adaptive and Penalized Estimation of Splines (SHAPES)~\citep{mohanty2020adaptive, mohanty2023}.

Moreover, we discuss the challenges inherent in glitch removal, including the trade-offs between sensitivity and specificity, computational complexity, and adaptability to evolving detector configurations. This review aims to provide a comprehensive overview of the current landscape of glitch removal techniques in gravitational wave data analysis by synthesizing insights from a wide range of literature spanning experimental observations, theoretical frameworks, and computational simulations.

Furthermore, we identify promising future research and development directions, including integrating machine learning approaches, refining data quality assessment criteria, and optimizing real-time glitch detection algorithms. Ultimately, the advancements in glitch removal systems discussed herein are poised to enhance the scientific yield of gravitational wave observatories significantly, enabling the detection of fainter signals, improving parameter estimation accuracy, and facilitating novel discoveries across the gravitational wave astronomy landscape. This review serves as a valuable resource for researchers and practitioners engaged in the ongoing quest to unlock the secrets of the universe through precision gravitational wave data analysis.

The rest of the paper organized as follows: Section~\ref{sec:Glitch} explain the glitch in the gravitational wave data, section~\ref{sec:Class_Glitch} briefly illustrates the glitch classification system, section~\ref{sec:Methods for removing glitches} reviews the background of the methods for glitch subtractions in the gravitational wave data, section~\ref{sec:Discussion} will explain the strength and limitations of the methods, section~\ref{sec:conclusion} will discuss the future work and concluding remarks on the discussion. 

\section{Glitch} 
\label{sec:Glitch}

Glitch is a type of non-stationary transient noise, which mimics the GW signal. In other words, a glitch is a non-GW signal, produced by non-astrophysical sources. As it mimics GW signals, it causes problems with the detection of GW signals. Additionally, it is difficult to increase the sensitivity of detectors due to the glitches~\citep{Buikema_2020, Powell_2018}. Moreover, the glitch results in inaccurate parameter estimation~\citep{Powell_2018}. The sensitivity of the GW detectors is increasing day by day. Thus, the probability of the number of glitches overlapping with GW signals in the GW data is rising with the progress of detectors. Despite the overlap of the glitches with GW signals, detectors are able to detect GW signals in the case of long-duration signals (for example GW170817), But for the short-duration GW signals, it will be difficult, in some cases not possible, to distinct glitches and GW signals. Therefore, removing glitches from GW data became very important.

A relatively recent approach to glitch subtraction is that of estimating the waveform of a glitch and subtracting it from the data. Every glitch has unique waveforms. In addition, the waveform of glitches changes slightly with the detector's design. As the glitches do not have known waveforms, it is difficult to detect, estimate and remove.


\section{Classification of glitches}
\label{sec:Class_Glitch}

Despite having different waveforms, glitches frequently belong to a number of different broad morphological classifications. The morphology of glitches in the time-frequency domain has been used to categorize glitches, which has helped researchers understand the origins of some classes and develop successful mitigation techniques for others. For the classification of glitches, a variety of methods—from fully automated to manual—have been suggested. The Gravity Spy project~\citep{Zevin_2017}] employs a citizen science strategy to include members of the general public in categorizing glitches through visual inspection of their Q-transform~\citep{Chatterji_2004} photographs.  As a result, several major glitch classes with descriptive names like Blip, Tomte, and Koi fish have been identified in the observation runs of the LIGO detectors thus far. Automated classification methods based on machine learning techniques have been proposed, including support vector machines~\citep{Biswas2013}, t-Sne~\citep{BAHAADINI2018}, random forests~\citep{Biswas2013}], S-means~\citep{Mukherjee_2010}, and deep convolutional neural networks~\citep{Biswas2013}. Two automated methods have been developed to detect errors using the Q-transform in addition to machine learning techniques: Omicron~\citep{ROBINET2021} and a method employing the p-value of the Q-transform~\citep{Leah_2022}. Changes in a detector's state, its couplings to the environment, and modifications to its hardware can affect the rates of glitches within a class as well as the emergence and disappearance of the classes themselves. 

\section{Methods for subtracting noises} 
\label{sec:Methods for removing glitches}

As removing glitches from GW data has paramount importance for detecting GW signals and increasing the sensitivity of detectors, various methods have been developed and applied to remove glitches. The following subsections contain a brief discussion of these techniques. 

\subsection{Wiener Filter and Regression with Wilson-Daubechies-Meyer (WDM) transformation} 

aLIGO has hundreds of auxiliary channels along with the GW channel to collect data. Some of the auxiliary channels record the data from the environment and some others collect data from instruments. Seismic noise is recorded using seismometers and accelerometers. Using a seismometer and accelerometer for seismic noise, the Wiener filter (the expectation value of the square of the error signal) can be built~\citep{Drigger_2012}. The error signal $\displaystyle{\left(\boldsymbol{e}_s\right)}$ can be defined as,
    \begin{align}
        {\bf e}_s &= {\bf n} - {\bf w}_y
    \end{align}
where $\displaystyle{{\bf n}}$ is the noise and $\displaystyle{{\bf w}_y} = {bf \omega}^T {\bf x}$ is the approximation of the noise, which can be found using auxiliary channels. Also, $\displaystyle{{\bf \omega}}$ is the tap weights of the filter, and $\displaystyle{{\bf x}}$ is the measurement of the external disturbance of the witness mirror. Therefore, the figure of merit for calculating the Wiener filter can be expressed as~\citep{Drigger_2012},
    \begin{align}
        E[{\bf e}_s^2] &= E[{\bf n}^2] - 2 {\bf \omega}^T p + {\bf \omega}^T R {\bf \omega}
    \end{align}
where $\displaystyle{p}$ is the cross-correlation vector between the witness channel and target noise, and $\displaystyle{R}$ represents the auto-correlation matrix for the witness channels. Additionally, an online adaptive filtering method (which is based on Least mean square methods) is applied, which gives the same result as the wiener filter~\citep{Drigger_2012}. This feed-forward method can be applied to higher frequency using an upgraded version of the filter~\citep{Tiwari_2015}, using data from auxiliary channels/ physical environment monitors (PEM). This method is built by modifying the Wiener-Kolmogorov (WK) filters~\citep{Weiner_1949} with regression in the time-frequency domain. The noise $\displaystyle{\left(n_w\right)}$ in the GW strain data $\displaystyle{\left(h\right)}$ can be predicted, using auxiliary channels $\displaystyle{\left(w\right)}$, as~\citep{Tiwari_2015}
    \begin{align}
        n\left[i\right] &= \sum_{j=-L}^{L} a_j w_{i+j}
    \end{align}
where the filter length is $\displaystyle{2L + 1}$, and $\displaystyle{a_j}$ is the filter coefficients. The filter coefficient can be determined by solving~\citep{Tiwari_2015}
    \begin{align}
        R_{xx} a &= p_{tx}
    \end{align}
which is Wiener-Hopf (WH) equation. Here, $\displaystyle{R_{xx}}$ is the auto-correlation matrix with $\displaystyle{\left(2L+1\right)\times \left(2L+1\right)}$ components, and $\displaystyle{p_{tx}}$ is the cross-correlation matrix in between the witness channels and the target noise with $\displaystyle{\left(2L+1\right)}$ components. Also, the filter can be obtained by 
    \begin{align}
        \chi^2 &= \sum_{i=L}^{N+L}\left(h[i] - \sum_{j=-L}^{L} a_j w_{i+j}\right)^2
        \label{eq:weiner_singlchannel}
    \end{align}
minimizing the mean square error. Regression with the WK filter has two major problems: WK filter requires long filters and computational complexity associated with the inversion of the matrix. Moreover, the filter may fail to capture all details of $\displaystyle{p_{tx}}$ due to spectral leakage. These problems can be solved by using Wilson-Daubechies-Meyer (WDM)~\citep{Necula_2012} transformation. But the regression with a single channel is not effective. Thus, multiple channels have been used to do the regression, and equation~\ref{eq:weiner_singlchannel} becomes
    \begin{align}
        \chi^2 &= \sum_{i=L}^{N+L}\left(h[i] - \sum_{j=-L}^{L} a_j w_{i+j} - \sum_{k=-L}^{L} b_k w_{i+k} - \sum_{m=-L}^{L} c_m w_{i+m} - \dots \right)^2 
    \end{align}
    
The introduction of multiple channel analysis illustrated two shortcomings: the matrix can be ranked deficient if auxiliary channels are highly correlated, and it can add noise if a significant fraction of the auxiliary channels are not correlated with target channels. Regulators have been introduced to solve this issue~\citep{Tiwari_2015}.


\subsection{Method for mitigating bilinear noise and scattering glitch using test mass} 
\label{subsec:Method for mitigating bilinear noise}

Bilinear noise can be mitigated through estimating bilinear noise by constructing a coherent bilinear noise filter using narrow-band noise in the signal recycling test mass and subtracting it from the data~\citep{Mukund_2020}. For improving the accuracy of the estimation, the adaptive Bayesian approach has been applied. Using the photodiodes in the end benches of VIRGO detectors, the scattered light glitch has been modeled and subtracted from the data~\citep{Was_2021}. When modeling, sine and cosine functions were utilized, however in actuality, the $tanh$ function was chosen since it produces better results and is more flexible. 

\subsection{\textit{Bayeswave}} 
\label{subsec:Bayeswave}

According to the assumption of the \textit{BayesWave}, the GW strain data $\displaystyle{\left[h(t)\right]}$ can be divided into three parts: these are Gaussian noise $\displaystyle{\left[n(t)\right]}$, signal $\displaystyle{\left[s(t)\right]}$, and glitch-transient noise $\displaystyle{\left[g(t)\right]}$. Mathematically, $\displaystyle{h(t)}$ can be written as,
    \begin{align}
        h\left(t\right) &= n\left(t\right) + s\left(t\right) + g\left(t\right)
    \end{align}

Using this notion, the \textit{BayesWave} algorithm estimates non-Gaussian features using Morlet-Gabor wavelets~\citep{Cornish_2015, Pankow_2018}, which is a sum of sine-Gaussian. The wavelets, for time series, can be written as
    \begin{align}
        \psi\left(t;\alpha \right) &= Ae^{-\frac{\left(t-t_0\right)^2}{\tau^2}} cos\left[2\pi f_0 \left(t-t_0\right) + \phi_0\right] 
    \end{align}
Here, $\displaystyle{\alpha = \left(A,t_0,q,f_0,\phi_0\right)}$ (amplitude, central time, quality factor $\left[q = 2 \pi f_0 \tau\right]$, central frequency, phase respectively) are the wavelet parameters. \textit{BayesWave} uses Bayesian inference to model non-stationary data, also the number and parameters of wavelets are not pre-determined. The number and parameter of wavelets were marginalized using Markov chain Monte Carlo (MCMC)~\citep{Cornish_2015}. Using the coherence in between detectors, the glitch and signal, non-stationary sources, can be determined. If the non-stationary source is coherent in between the detectors, than the model for this part considered as signal $\displaystyle{\left[s(t)\right]}$ while the incoherent part is considered as glitch $\displaystyle{\left[g(t)\right]}$~\citep{Pankow_2018}. In the case of a single detector signal, the glitch modeling from auxiliary channels can be used to remove the glitches around the signal~\citep{Davis2022}.

\subsection{\textit{gwsubtract}} 
\label{subsec:gwsubtract}

A linear subtraction method, \textit{gwsubtract}, is applied to remove the glitch~\citep{Davis_2019}. It uses auxiliary channels to determine transfer function $\displaystyle{\left[T_f\right]}$. Using this transfer function into the strain data, it estimates a glitch and other noises. After that, the glitch is removed by subtracting the estimate from the data. In case of \textit{gwsubtract} method, the assumption is that the linear combination of timeseries from various sources $\displaystyle{\left[n_j (t)\right]}$ produce the GW strain data $\displaystyle{\left[h(t)\right]}$. Or the assumption can be considered as $\displaystyle{h(t)}$ consists a signal $\displaystyle{\left[s(t)\right]}$, which is not correlated with the noise~\citep{Allen1999}, in the GW strain data. Also, another important assumption is that the multiplication of the convolution of a witness time series $\displaystyle{\left[w(t)\right]}$ and an unknown transfer function $\displaystyle{\left[T_{fwh}\left(t\right)\right]}$ can model the noise, at least one of the noises~\citep{Allen1999,Davis2022}. Therefore, 
$\displaystyle{h(t)}$ can be written as,
    \begin{align}
        h\left(t\right) &= n_1\left(t\right) + n_2\left(t\right) + n_3\left(t\right) + \dots + n_{j-1}\left(t\right) + w\left(t\right) \times T_{fwh}\left(t\right) \nonumber\\
        &= h^{'}\left(t\right) + w\left(t\right) \times T_{fwh}\left(t\right)
        \label{eq:gwsubtract_strain1}
    \end{align}
    
In equation~\ref{eq:gwsubtract_strain1}, $\displaystyle{h^{'}\left(t\right)}$ has all of the property of strain data, except one noise sources modeled by the $\displaystyle{w\left(t\right)}$ and the $\displaystyle{T_{fwh}\left(t\right)}$. In frequency domain, the time domain equation~\ref{eq:gwsubtract_strain1} can be expressed as
    \begin{align}
       \Tilde{h}\left(f\right) &= \Tilde{h}^{'}\left(f\right) + \Tilde{w}\left(f\right) \times \Tilde{T}_{fwh}\left(f\right)
    \end{align}
    
From the derivation from~\citep{Allen1999,Davis_2019}, the transfer function $\displaystyle{\Tilde{T}_{fwh}\left(f\right)}$ for frequency band $\displaystyle{\left[f_{j-1}, f_j\right]}$ can be written as,
    \begin{align}
        \Tilde{T}_{fwh}\left(f'\right) &= \frac{df}{f_j - f_{j-1}}  \sum_{f_{j-1}}^{f_j} \Tilde{Y}_h\left(f\right) \Tilde{Y}_w^*\left(f\right) 
    \end{align}
where $\displaystyle{\Tilde{Y}_h\left(f\right)}$ is the discrete Fourier transform of $\displaystyle{h(t)}$, $\displaystyle{\Tilde{Y}_w^*\left(f\right)}$ is the discrete Fourier transform of $\displaystyle{w(t)}$, and $\displaystyle{df}$ is the frequency resolution of the data. The chance correlations measurement has been reduced by averaging nearby frequencies during the transfer function calculation~\citep{Davis2022,Allen1999}. Note that the inner product in the transfer function equation can be determined if $\displaystyle{h\left(f\right)}$ and $\displaystyle{w\left(f\right)}$ are sampled quickly.

\subsection{\textit{glitschen}} 
\label{subsec:Glitschen}
\textit{glitschen} is a data-driven, parametric glitch mitigation model~\citep{Merritt2021}, which uses probabilistic principal component analysis (PPCA) method~\citep{Tipping1999}. In \textit{glitschen}, an isotropic Gaussian noise model with a $\displaystyle{d-}$ dimensional observation vector $\displaystyle{\left(\Tilde{\boldsymbol{d}}\right)}$ can be expressed as,
    \begin{align}
        \Tilde{{\bf d}}|{\bf X}_{t} \sim N\left({\bf W} {\bf X}_{t} + \mu, \sigma^2 {\bf I}\right)
    \end{align}
where the mean of isotropic Gaussian noise $\displaystyle{\left(\mu = 0\right)}$is replaced by $\displaystyle{{\bf W} {\bf X}_{t} + \mu}$, $\displaystyle{\sigma^2}$ is variance, and $\displaystyle{{\bf I}}$ is the identity matrix. Also, $\displaystyle{{\bf X}_{t}} \sim N\left(0, {\bf I}\right)$ is marginalized by $\displaystyle{r}$ latent variables of the training set, and $\displaystyle{{\bf W}}$ has dimensions $\displaystyle{d \times r}$ with $\displaystyle{r}$ training eigenvectors.

Following the derivation of~\citep{Merritt2021, Tipping1999}, a new observation vector $\displaystyle{\left(\Tilde{{\bf d}}_o\right)}$ can be written as
    \begin{align}
        {\bf X}_{t}|\Tilde{{\bf d}}_0 \sim N\left({\bf M}^{-1} {\bf W}^T \left(\Tilde{{\bf d}}_o - \mu \right), \sigma^2 {\bf M}^{-1}\right)
    \end{align}
where $\displaystyle{{\bf M} = {\bf W}^T {\bf W} + \sigma^2 {\bf I}}$, with size $\displaystyle{r \times r}$.

Using this observation vector, the glitch can be modeled by
    \begin{align}
        \Tilde{{\bf g}}_m = {\bf W} {\bf X}_m + \mu
    \end{align}
where $\displaystyle{{\bf X}_m \equiv {\bf M}^{-1} {\bf W}^T \left(\Tilde{{\bf d}}_o - \mu \right)}$. The quality of the glitch model is determined by the standard Gaussian noise likelihood.

\subsection{\textit{DeepClean}} 
\label{subsec:DeepClean}

In glitch subtraction methods, a convolutional neural network is used. \textit{DeepClean} is such a noise removal method~\citep{Ormiston_2020}, which uses deep learning and one dimensional convolutional neural network (CNN). For this method, the GW strain data $\displaystyle{\left[h\left(t\right)\right]}$ is considered to be the combination of fundamental noise $\displaystyle{\left[n_f\left(t\right)\right]}$, signal $\displaystyle{\left[s\left(t\right)\right]}$, and other noise $\displaystyle{\left[n_w\left(t\right)\right]}$. Therefore, $\displaystyle{h\left(t\right)}$ can be expressed as
    \begin{align}
        h\left(t\right) &= s\left(t\right) + n_f\left(t\right) + n_w\left(t\right)
    \end{align}
Here, $\displaystyle{n_f\left(t\right)}$ is not desired to be subtracted, while $\displaystyle{n_w\left(t\right)}$, which couples into the witness/auxiliary channels $\displaystyle{\left[w_j\left(t\right)\right]}$, is needed to be subtracted. In this method, the data from auxiliary channels, after preprocessing, is sent to CNN. For estimating $\displaystyle{n_w\left(t\right)}$, CNN work as a function $\displaystyle{\mathcal{F}\left(w_j\left(t\right);\overrightarrow{\beta}\right)}$ on $\displaystyle{w_j\left(t\right)}$. Here, $\displaystyle{\overrightarrow{\beta}}$ is a set of parameters,that can be determined by
    \begin{align}
        \overrightarrow{\beta} &= argmin_{\overrightarrow{\beta}'} \mathcal{J}\left[h\left(t\right), \mathcal{F}\left(w_j\left(t\right);\overrightarrow{\beta}'\right)\right]
    \end{align}
where $\displaystyle{\mathcal{J}}$ is a loss function, which is determined by the sum of the loss function $\displaystyle{\mathcal{J}_{asd}}$ (the weighted average of the amplitude spectral density (ASD) of the residual strain $\displaystyle{\left[r\left(t\right)\right]}$ [where $\displaystyle{\left[r\left(t\right) = h\left(t\right) - \mathcal{F}\left(w_j\left(t\right);\overrightarrow{\beta}\right)\right]}$] and the time domain loss function $\displaystyle{\mathcal{J}_{mse}}$ (the mean square error (MSE) across the time series). Therefore, $\displaystyle{\mathcal{J}}$ can be written as,
    \begin{align}
        \mathcal{J} &= m \mathcal{J}_{asd} + \left(1-m\right) \mathcal{J}_{mse}
    \end{align}
where $\displaystyle{m}$ is a weighting factor ranging from $\displaystyle{0}$ to $\displaystyle{1}$.

Using this similar method, a CNN method has been built to remove glitches~\citep{Magoushi_2021}. The difference between \textit{DeepClean} and this CNN method is that it has two-dimensional CNN. The non-linear activation, \textit{ReLU}, has been used for this CNN method.

\subsection{Method for mitigating angular noise} 
\label{subsubsec:Method for mitigating angular noise}

A CNN method~\citep{yu2021} has been built for mitigating angular noise using the code \textit{Keras}~\citep{chollet2018}(a deep learning code written in python), which is running on \textit{TensorFlow}~\citep{Abadi2015}. Angular noise is caused by a geometrical effect of the rotational pivot of aLIGO test mass. Also this nonlinear noise couples with the GW readout. The angular motion in the test mass can be expressed as~\citep{yu2021}, 
    \begin{align}
        \delta x^{\left(mir\right)}\left(t\right) &=  x_{spot}^{\left(mir\right)}\left(t\right) \theta^{\left(mir\right)}\left(t\right)
    \end{align}
where $\displaystyle{\delta x^{\left(mir\right)}\left(t\right)}$ is a linear length fluctuation, $\displaystyle{x_{spot}^{\left(mir\right)}\left(t\right)}$ is a slow $\displaystyle{\left(\lesssim 1\ Hz \right)}$ motion of the beam spot on the test mass, and $\displaystyle{\theta^{\left(mir\right)}\left(t\right)}$ is a fast $\displaystyle{\left(\gtrsim 10\ Hz \right)}$ motion of the beam spot on the test mass. The slow motion induced by the seismic motion and the fast motion, angular perturbation of the mirror, induced by the sensing noise in the angular control system~\citep{yu2021}.

In this method, there are two types of CNN structure: General CNN and Specific CNN. General CNN consists of sufficiently many convolutional layers. These layers are densely connected and at least some of these layers have a nonlinear activation function. For the layers with nonlinear activations, the coupling mechanism can be expressed with series expansion and the convolutional layer will act as a finite-impulse-response (FIR) filter. However, using the knowledge of angular noise, a more specific CNN structure can be built, which consists of the auxiliary channels of a slow CNN and the auxiliary channels of a fast CNN. These sets, slow and fast, require only linear activation. This method also works as an FIR filter to convert the linear outputs of auxiliary channels into digital counts. The nonlinearity of the noise can be determined by using equation $\displaystyle{\left(2.12\right)}$, i.e. by multiplying layers. In both CNN structures, a loss function has been applied which is similar to the loss function of \textit{DeepClean}. For getting better noise mitigation, CL training and techniques~\citep{GEORGE2018} can be used with this method.
\subsection{SHAPES} 
\label{subsec:SHAPES}

The fundamental algorithm behind SHAPES~\citep{mohanty2020adaptive}, is derived using the following models for the signal and noisy data, $\overline{s}(\theta)$ and $\overline{y}$, respectively. 
\begin{eqnarray}
\overline{y} & = & \overline{s}(\theta) + \overline{\epsilon}\;,
\label{eq:regressionModel}
\end{eqnarray}
where $y_i = y(t_i)$ and $s_i(\theta) = s(t_i;\theta)$, $i = 0,1,\ldots,N-1$, are sampled values at $t_i$ with $t_0 = 0$, $t_{N-1} = 1$, $t_{i+1}>t_i$, and $\theta$ signifies the set of signal parameters that need to be estimated from the data, $\overline{y}$, $\overline{s}$, and $\overline{\epsilon}$ are row vectors with $N$ elements.
The noise samples, $\epsilon_i$, are randomly selected from the normal (Gaussian) probability density function with zero mean and unit variance, $N(0,1)$. Since GW data is always whitened using the estimated noise power spectral density (PSD), this assumption—that of a white Gaussian noise process—does not result in a loss of generalization. 

Since it is assumed that the signal $s(t;\theta)$ is a spline of order $k$, it can be represented by a linear combination of B-spline functions~\citep{deBoor2001}. 
\begin{eqnarray}
s\left(t;\theta = \{\overline{\alpha},\overline{\tau}\}\right) &=& \sum_{j = 0}^{P-k-1}\alpha_j B_{j,k}(t;\overline{\tau})\;,
\label{eq:fittingFunction}
\end{eqnarray}
where $\overline{\alpha}= (\alpha_0,\alpha_1,\ldots,\alpha_{P-k-1})$, and $\overline{\tau}= (\tau_0, \tau_1, \ldots, \tau_{p-1})$, $tau_{i+1}\geq \tau_i$ is a series of $P$ knots that identifies the end points of the continuous intervals holding the cubic polynomial components composing the spline. The set of B-spline functions, $\{B_{i,k}(x;\overline{\tau})\}$, for any given order can be efficiently computed using the Cox-de Boor recursion relations in~\citep{DEBOOR197250}.

$\widehat{\alpha}$ and $\widehat{\tau}$ are the spline parameters that minimize a penalized least-squares function, and they provide the best fit, 
 \begin{eqnarray}
L_\lambda(\overline{\alpha},\overline{\tau}) & = & L(\overline{\alpha},\overline{\tau}) + \lambda R(\overline{\alpha})\;,
\label{eq:penalizedSpline}\\
L(\overline{\alpha},\overline{\tau}) & = & \sum_{i = 0}^{N-1} \left(y_i - s_i(\overline{\alpha},\overline{\tau})\right)^2\;,
\end{eqnarray}
where the penalty term, $$R(\overline{\alpha}) = \sum_{j=0}^{P-k-1} \alpha_j^2\;$$,
is discovered to be effective in the inhibition of erroneous knot clustering. When the best-fit spline attempts to reduce $L_\lambda(\overline{\alpha}, \overline{\tau})$ by fitting out anomalous data points caused by noise alone, clusters like these are seen. In the present version of SHAPES, the penalty gain factor, $\lambda$, is user-specified: a greater value of $\lambda$ forces the best suited spline to be smoother. Given that they exist linearly in the signal model, optimizing over $L_\lambda(\overline{\alpha}, \overline{\tau})$ over $\overline{\alpha}$ is simple. However, optimizing over $\overline{\tau}$ has long been a hurdle~\citep{wold1974spline,burchard1974splines,jupp1978approximation,luo1997hybrid} when implementing adaptive spline fitting. The advantages of optimizing knot location have also been amply established at the same time. This has inspired a variety of approaches for the knot optimization step in the literature.  It is evident from~\citep{galvez2011efficient, mohanty2012particle} that Particle Swarm Optimization (PSO)~\citep{PSO,mohanty2018swarm}, a well-liked nature-inspired metaheuristic for global optimization of high-dimensional non-linear and non-convex functions, provides for significant advancement on this topic.

\section{Discussion on the methods} 
\label{sec:Discussion}

Feed-forward methods can increase the sensitivity of aLIGO detectors and enhance several percent of detectable inspiral range~\citep{Meadors_2014}. Wiener filter, a feed-forward method, is applied to only seismic noise below $\displaystyle{10}$ Hz; more specifically, it was successful in reducing noise in between $\displaystyle{0-7}$ Hz~\citep{Drigger_2012}. Feed-forward methods can be applied up to $\displaystyle{150}$ Hz, and using auxiliary channels also increases the sensitivity of the detectors~\citep{Tiwari_2015}. The system, built with the Wiener filter, is applied to reduce the seismic noise below $\displaystyle{10}$ Hz. It cannot be applied above $\displaystyle{10}$ Hz as it can remove gravitational waves along with the noise. From the results, it indeed reduced the seismic noise and smoothed the data, but it can not smooth all the parts. Moreover, it is applied at a very low frequency. Also, narrow-band and broadband noises need to be removed more successfully. The method, built with regression, is also successful in removing bilinear noise produced by linear coupling of seismic noise and resides near the power lines and calibration lines~\citep{Tiwari_2015}. The nonlinear cases are a challenge for this system. They had to input the data from multiple channels to confirm the estimation. In this case, it overfits the data, and the system has a problem with rank deficiency of linear equations. Also, it is applied for only low-frequency noise, below $\displaystyle{150}$ Hz to be exact. The regulator solves the problem related to multiple-channel analysis. However, the value of regulators can not be generalized. The value of the regulator and the regression parameters should be changed with the noise structure.

The bilinear noise removal method depends on how well the system makes the model of narrow-band noise for the success of this method. It works on bilinear noise~\citep{Mukund_2020}, but it is still being determined whether it will work on glitches as glitch has various waveforms. The scattered glitch that has been mitigated is in between $\displaystyle{30-55}$ Hz~\citep{Was_2021}]. If a scattered light comes in a higher frequency, it can not be obliterated. Also, the effect of this mitigation on the GW signal is not explained.

The success of the glitch removal, using \textit{BayesWave}, depends on how well the signal is modeled~\citep{Cornish_2015}. The coherence method~\citep{Pankow_2018} can not model a glitch when the signal is only detected in one detector (for example, GW190424). For the success of the glitch removal method from a single detector, an auxiliary channel should model the glitch very well, or it will affect the signal. \textit{BayesWave} can not model a glitch accurately if the glitch duration is  $\displaystyle{ > 1}$s~\citep{Davis2022}. Hence, it can not accurately model scattering glitches~\citep{Davis2022} or any other glitch that is longer than $\displaystyle{1}$s. Although \textit{BayesWave} can model a glitch more accurately when the glitch duration is $\displaystyle{\lesssim 1}$s, it has an error percentage depending on the glitch model from auxiliary channels and may bias the result slightly.

The GW strain data are nonstationary and non-Gaussian due to noise artifacts of various durations ~\citep{Abbott2018}. Therefore, \textit{gwsubtract} is not suitable for removing glitches in maximum cases. If the glitch system is linear, \textit{gwsubtract} can remove the glitches~\citep{Davis2022}. Nevertheless, in this case, it will depend on the noise model made by the auxiliary channels. Though the statistical error and systematic uncertainties from \textit{BayesWave} and \textit{gwsubtract} are low~\citep{Hourihane2022}, it has a more considerable impact on the measured source properties~\citep{Pyne2022}.

The method explained in \textit{glitschen} is fast, and the training for this method is computationally cheap. The method can successfully model glitches (Blip, Tomte) and subtract them from the data. Also, the method can sometimes differ between astrophysical signals and glitches. The problem is that the method requires a lot of training data and training model~\citep{Merritt2021}. The glitch modeling depends on how high the SNR of the glitch in the data is~\citep{Merritt2021}. Also, as the sensitivity of aLIGO increases, the glitches' waveforms are changing. So, the method will require identifying some glitches in the selected observing run first and then training the system to estimate and remove the glitches. Moreover, the method can not differentiate between an unknown astrophysical signal and a new glitch, which may cause confusion in the system and can remove the astrophysical signals. \textit{glitschen} performs better in the case of tomte than a blip. The tomte glitch that has been described in the work~\citep{Merritt2021} has a very low frequency. The method may work better below 100 Hz, but above that, this system does not perform at its best. \textit{BayesWave} is also applied for modeling the glitches and the astrophysical signals~\citep{Chatziioannou_2021}. \textit{BayesWave} can successfully distinguish the glitches and astrophysical signals in the case of binary black hole (BBH) signals, but the system needs to be trained with various signals before applying it~\citep{Chatziioannou_2021}. This is a problem as many astrophysical signals are still not modeled and known. Therefore, the performance of \textit{BayesWave} depends on how well the method is trained. That means \textit{BayesWave} needs a plethora of data to train itself first. Also, the limitations previously discussed persist in this case as well.

\textit{DeepClean} is mainly applied to remove seismic noise, jitter noise, and sidebands. It works as a Weiner filter but also has some advantages. It is not strictly used for linear couplings. Instead, it uses the $\displaystyle{tanh}$ function, and thus, it can discern some nonlinear features. One of the main advantages of \textit{DeepClean} is that it can determine linear, nonlinear, and nonstationary couplings without any prior knowledge of these noises, i.e., without prior training of the particular noise type~\citep{Ormiston_2020}. Though a fraction of the scattering glitch can be removed using the dimensional CNN method, it cannot obliterate it or the maximum part of the glitch~\citep{Magoushi_2021}. The method performs poorly on Extremely loud glitches as scattering glitches~\citep{Magoushi_2021}. That means the CNN method estimation is a small fraction of the extremely loud glitch. One of the positive things about this method is that this method does not take out too much power from the data. The system uses auxiliary/witness channels for training. However, adding auxiliary channels, which are irrelevant in case of a particular glitch, can cause uncertainty to the subtraction~\citep{Ormiston_2020, Magoushi_2021}. Hence, selecting auxiliary channels for training is crucial for these methods.
Moreover, the length of training data depends on the complexity of the particular noises. In general, CNN needs a large amount of data to train itself. Also, the computational cost for CNN techniques is relatively high.

For the simple structure of general CNN in the method for angular noise mitigation, it needs a slight knowledge of noises to mitigate noise even with unknown couplings~\citep{yu2021}. Conversely, detailed knowledge of noise is needed to build specific CNN. Also, the challenging part for specific CNN is the reconstruction of the spot motion $\displaystyle{x_{spot}^{\left(mir\right)}\left(t\right)}$ on test mass. The performance of general CNN and specific CNN in noise subtraction is decent and comparable in between $\displaystyle{10-20}$ Hz~\citep{yu2021}. However, the performance is better for specific CNN between $\displaystyle{6-10}$ Hz. Also, after $\displaystyle{30}$ Hz, which is the end of the training band, general CNN starts to add noise while specific CNN continues to remove noises from data. On the real aLIGO data, the broadband reduction of nonlinearity is not that significant~\citep{yu2021}. Another significant condition for these methods is the signal-to-noise ratio (SNR) value. A slight change in the SNR has negatively affected the noise subtraction significantly.

\textit{SHAPES} uses non-parametric regression analysis method~\citep{mohanty2020adaptive,cruz2021data,mohanty2023}, unlike the other methods, that are discussed in section~\ref{sec:Methods for removing glitches}. One of the privilege of this method is that \textit{SHAPES} do not need any previous knowledge of glitches to subtract it from the GW data. It works on various types of glitches without knowing the type of it~\citep{mohanty2023}. However, the process is not fully automatic yet~\citep{mohanty2023}. \textit{SHAPES} perform well when the boundaries of glitch have been identified well enough~\citep{Chowdhury2022-wf}. Hence, the boundaries of glitches should be identified accurately in this case. Thus, there is a lot of work that needs to be done to identify the glitch boundary more accurately for \textit{SHAPES} best performance. About the time and computational cost to run the code, \textit{SHAPES} takes much less time to remove the glitch than \textit{Bayeswave}. Consequently, \textit{SHAPES} will need less computational cost then \textit{Bayeswave}.

From the discussion above, the methods explained above are effective on linear noises in various frequencies. In aLIGO, the linear component of noises has been mitigated successfully~\citep{Davis_2019,Drigger_2019}. However, in case of glitches, \textit{BayesWave}, \textit{gwsubtract} and a two dimensional CNN structure similar to \textit{DeepClean} can be applied. But none of these methods can work without the model of glitches from auxiliary channels. Also, in some cases, auxiliary channels can not catch glitch waveform accurately and in these cases, these methods will fail. It requires a lot of training to train the model for CNN base methods. \textit{gwsubtract} can not remove glitches which is not linearly dependent. Though \textit{glitschen} is successful for removing glitches at a certain level, it requires prior knowledge of glitches to remove the glitch effectively. Moreover, it has uncertainties for estimating and mitigating the wing of the glitches. \textit{SHAPES} can subtract the glitches from GW data without having any prior knowledge about the glitches, however the system is not automatic and need human to define the glitch boundary.

\section{Conclusions}
\label{sec:conclusion}

For Advanced LIGO (aLIGO) detectors to function more accurately and sensitively, efficient glitch subtraction techniques are essential. While approaches such as feed-forward and regression-based techniques have demonstrated potential in mitigating low-frequency seismic and bilinear noise, difficulties arise when dealing with higher frequencies and nonlinear scenarios. Convolutional neural networks (CNNs), \textit{BayesWave}, \textit{DeepClean}, and other advanced techniques provide better handling of complicated noise, but they demand a large amount of computer power and training. Although it is not entirely automated and depends on precisely defined glitch boundaries, \textit{SHAPES} stands out for its capacity to eliminate a variety of glitches without prior knowledge. Although these techniques work well for linear sounds, precise auxiliary channel models are frequently necessary for these techniques to work well for glitches. Subsequent endeavors have to concentrate on hybrid methodologies, enhanced automated identification, and flexibility in response to changing noise attributes.

\bibliographystyle{plainnat} 
\bibliography{referances}






\end{document}